\documentclass[prb,showpacs,twocolumn,superscriptaddress]{revtex4}

\usepackage{amsmath,graphicx,latexsym,bm}
\usepackage{epstopdf}

\begin{document}

\title{Electrical properties of improper ferroelectrics from first principles}

\author{Massimiliano Stengel}
\affiliation{ICREA - Instituci\'o Catalana de Recerca i Estudis Avan\c{c}ats, 08010 Barcelona, Spain}
\affiliation{Institut de Ci\`encia de Materials de Barcelona 
(ICMAB-CSIC), Campus UAB, 08193 Bellaterra, Spain}

\author{Craig J. Fennie}
\affiliation{School of Applied \& Engineering Physics, Cornell University, Ithaca, NY 14853 US}

\author{Philippe Ghosez}
\affiliation{Physique Th\'eorique des Mat\'eriaux, University of Li\`ege, 4000 Li\`ege, Belgium}

\begin{abstract}
We study the interplay of structural and polar distortions in
hexagonal YMnO$_3$ and short-period PbTiO$_3$/SrTiO$_3$ superlattices
by means of first-principles calculations at constrained electric 
displacement field $D$.
We find that in YMnO$_3$ the tilts of the oxygen polyhedra produce
a robustly polar ground state, which persists at any choice of the
electrical boundary conditions. Conversely, in PTO/STO the 
antiferrodistortive instabilities alone do not break inversion symmetry, 
and open-circuit bundary conditions restore a non-polar state.
We suggest that this qualitative difference naturally provides a route
to rationalizing the concept of ``improper ferroelectricity'' from the
point of view of first-principles theory.
We discuss the implications of our arguments for the design of novel
multiferroic materials with enhanced functionalities, and for the 
symmetry analysis of the phase transitions.
\end{abstract}


\maketitle


\section{Introduction}

Multifunctional oxide materials, characterized by coexisting ferroic 
orders of different physical origin, are currently the object of 
intense exploration.
%
Particularly enticing for their potential technological applications 
and fundamental interest are the so-called multiferroic materials,
where spontaneous magnetic and ferroelectric orders coexist in 
the same phase.
%
%
%
Unfortunately, few multiferroic oxides are known, and their 
magnetoelectric coupling is usually too small (or occurs at 
exceedingly lower temperatures) for practical purposes. 
This is typically explained as follows: in the most popular ferroelectric oxides, 
such as BaTiO$_3$, the spontaneous polarization emerges from the hybridization
between occupied O($2p$) orbitals and empty $3d$ orbital of the transition-metal
cation.
This $d^0$-ness condition typically rules out magnetism. Although exceptions 
have been identified~\cite{Bhattacharjee:2009,Gunter:2012},
ferroelectricity and magnetism often appear to be \emph{contraindicated}.~\cite{Hill:2000}

One way to achieve a strong coupling between electrical and magnetic degrees
of freedom, is  to resort to a more restrictive class of materials, called 
``improper ferroelectrics'', where the polarization is induced by completely 
different microscopic mechanisms.~\cite{levanyuk} 
(Another less restrictive mechanism involves the idea of phase competition~\cite{fennie06,tokura07}.)
Strictly speaking, the distinction between proper and improper ferroelectrics is
made on symmetry grounds, and is based on the dynamics of the phase transition 
from a high-temperature centrosymmetric to a lower-temperature polar phase.
In proper ferroelectrics (PF), such as PbTiO$_3$ or BaTiO$_3$, the polar distortion 
acts as the primary order parameter in the phase transition; conversely, in improper 
ferroelectrics (IF) the polarization $P$ is slave to another, non-polar, order 
parameter of different physical origin through an energy term linear in $P$.
%
Depending on the material these can be the rotation/tilts of the oxygen 
polyhedra network, or the magnetic order associated, e.g. with spin cycloids.
Unlike covalency-driven polar distortions, these alternate mechanisms can potentially 
lead to strong magnetoelectric couplings. 
%
For example spin-driven improper ferroelectricity in TbMnO$_3$~\cite{mostovoy} naturally 
couples magnetic and polar degrees of freedom, resulting in a remarkable magnetoelectric
effect, while structurally-driven improper ferroelectricity in YMnO$_3$~\cite{Fennie/Rabe:2005} 
leads to an unusual locking of polar and antiphase structural domain walls~\cite{Cheong-10}.
These recent discoveries suggest an opportunity to achieve entirely new classes of strongly 
coupled multifunctional materials with novel properties based on improper ferroelectricity.
As we said above, however, improper ferroelectrics are not abundant in nature. 
As a result material design rules have remained elusive and the discovery
of new multiferroics occurs serendipitously.

This situation is now rapidly changing thanks to the increasing popularity of density-functional
theory calculations, which have led the quest for new oxide materials in the past few years.
An important breakthrough in the field~\cite{Bousquet/Ghosez:2008} occurred by the recent 
discovery of a new type of improper ferroelectricity in artificially layered PbTiO$_3$/SrTiO$_3$ (PTO/STO) 
superlattices. This system displays an unusual form of lattice-driven instability, involving 
the interplay of $P$ with octahedra rotations and tilts. More specifically, the free energy 
of this system  was shown to display an unusual trilinear coupling, $\sim$$\Phi_1 \Phi_2 P$, 
between the polarization and \emph{two} non-polar antiferrodistortive (AFD) modes of 
different symmetry, $\Phi_1$ and $\Phi_2$.  
This work has paved the way towards the rational design of lattice-driven improper 
ferroelectrics~\cite{Rondo:2012} with a large spontaneous $P$ and has lead to the discovery of a 
novel class of (CaMnO$_3$)$_2$CaO strongly coupled multiferroics~\cite{Benedek/Fennie:2010}. 
Given the versatility of perovskites, this novel class of IFs are ripe systems to explore for 
new multifunctional  phenomena.

There are however important questions that still need to be answered from the point of view of 
fundamental theory. For example, the origin of the polarization is not always clear if in addition 
to the two AFD instabilities -- which from the trilinear coupling could lead to a $P\sim\Phi_1\Phi_2$ 
-- a proper ferroelectric instability exists. 
This issue is ultimately tied to the question of which two order parameters can be identified 
as  {\it inducing} the distorted state (e.g., in this case  $\Phi_1 \sim \Phi_2 P$ is an equally 
valid statement). This question of primary order parameters determines to what extent the 
electrical and dielectric properties behave like conventional IFs. 
%
This is of central importance for the applicability of  multifunctional materials in actual 
devices. 
Unfortunately the ability to clearly address this using symmetry arguments and/or 
conventional first-principles calculations alone is difficult.
What is clearly missing is a generally applicable and reliable ``test'' within a zero-temperature 
calculation, which would unambiguously  identify whether the properties of this class of (hybrid) 
IFs  behaved more like a conventional IF or more like a proper ferroelectric.

Here we compare from first principles the electrical properties of hexagonal YMnO$_3$ to 
that of the PTO/STO superlattice. We illustrate the key fingerprints of improper ferroelectricity 
by using a recent advance in first-principles methods~\cite{fixedd} that treats the electrical 
displacement, $D$, as the fundamental variable. This allows us to directly compute the spontaneous
polarization in open-circuit, $P_{\rm OC}$, within a periodic bulk calculation, avoiding the need 
for cumbersome slab geometries (where the results inevitably depend on the details of the surfaces; 
a common source of artifacts in the calculation).
%
We identify the presence of a nonzero and switchable $P_{\rm OC}$ as a necessary 
(although generally not sufficient) condition that must be satisfied by a material in order 
for it to be classified as improper ferroelectric. This property is enjoyed by YMnO$_3$ 
(\emph{albeit} with a substantially smaller spontaneous $P_{\rm OC}$ than reported 
previously), but not by the 1/1 PbTiO$_3$/SrTiO$_3$ superlattices considered here. 
We rationalize this unexpected result in 
terms of higher order couplings between the AFD mode(s) and the polarization. 
We briefly discuss the implications of these result for the
future design of novel improper ferroelectric materials.
This approach puts the concept of improper ferroelectricity on firmer conceptual 
grounds from the point of view of \emph{ab-initio} theory, providing a useful tool 
for the rational design of new ``hybrid''~\cite{Benedek/Fennie:2010} IFs.

\section{Computational methods}

%
Our calculations are performed within the local-density approximation
(LDA) of density-functional theory and the projector-augmented-wave (PAW)
method \cite{Bloechl:1994}, with a plane-wave cutoff of 60 Ry.
For YMnO$_3$, we apply a Hubbard correction (LDA+U) of U=6 eV to 
the Mn $3d$ states, in order to obtain an insulating ground state.
We further assume a collinear antiferromagnetic configuration of the
spins, and use a 4x4x2 Monkhorst-Pack~\cite{Monkhorst/Pack:1976} (MP) sampling 
of the 30-atom hexagonal unit cell to perform the Brillouin-zone integrals. 
We use standard non-spin polarized LDA for PTO/STO, together
with a 4x4x3 MP grid to sample the Brillouin zone
of the $\sqrt{2} \times \sqrt{2} \times 2$ direct-space cell
(20 atoms total, doubled in-plane to accommodate the 
antiferrodistortive rotations of the oxygen octahedra).
Cell relaxations are performed by using the calculated stress tensor,
which we correct for the Pulay error by applying a constant negative
pressure. For YMnO$_3$, first we optimize the structure of the ferroelectric 
phase, and then fix the in-plane lattice parameter to the equilibrium 
value in zero applied field. 
For PTO/STO we fix the in-plane lattice parameter to the 
equilibrium theoretical value of cubic SrTiO$_3$, as it has became 
common practice in simulations of layered perovskites. (We relax only
the out-of-plane lattice parameter in response to the electrical
perturbations, as typically these materials are grown as epitaxial 
films on a thick substrate, which prevents their in-plane relaxation.)
The macroscopic electrical degrees of freedom are dealt with
by using the methods described in Ref.~\onlinecite{fixedd}. In all cases we 
enforce rather stringent threshold for structure relaxations, as
the polarization and the internal electric fields are both very 
sensitive to minimal displacements of the ions.
More specifically, we calculate the ground state of both materials
as a function of the macroscopic electric displacement field,
and analyze the results in terms of energy, internal electric field,
amplitude of the relevant structural (zone-boundary) modes and 
out-of-plane (electrically induced) strain.

\section{Results}

\begin{figure}
\includegraphics[width=3.3in]{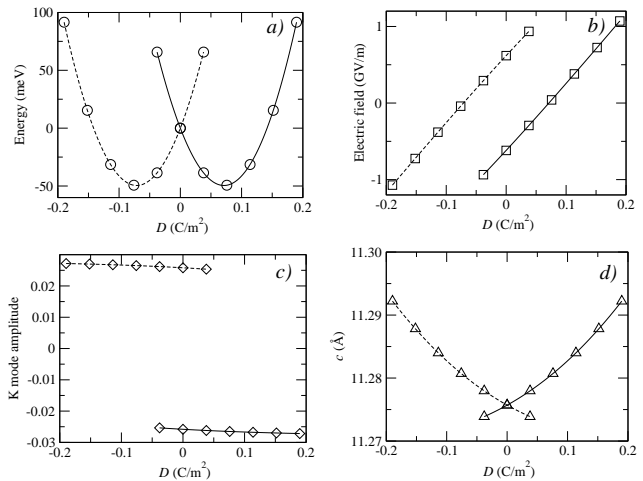}
\caption{Electrical properties of YMnO$_3$. The energy (a), internal
electric field (b), $K_3$ mode amplitude (c) and out-of-plane strain (d) are 
plotted as a function of the macroscopic $D$ field. Symbols refer to
first-principles data; the continuous lines are SPLINE fits to the symbols.}
\label{fig1}
\end{figure}

\subsection{YMnO$_3$}

%
In a proper ferroelectric, open-circuit boundary conditions suppresses single-domain ferroelectricity 
and drives domain formation. Imposing open-circuit boundary conditions corresponds to adding to the
free energy a positive term that is quadratic in $P$~\cite{fixedd},
\begin{equation}
\Delta E = \frac{\Omega}{8\pi} (D-4\pi P)^2,
\end{equation}
where $\Omega$ is the volume and $D$ is the electric displacement of the
high-temperature paraelectric phase, which can be taken as zero.
Remarkably, Sai et al.~\cite{Sai_et_al:2009} demonstrated by means of 
first-principles calculations that a monodomain state with non-zero $P$ can 
persist in YMnO$_3$ even in the extreme limit of an unsupported film without 
electrodes. 
Because the polarization couples linearly to the primary order parameter in an 
improper ferroelectric~\cite{levanyuk}, \emph{any} IF must enjoy such a property.
%

So are there clearly identifiable  features in the electrical diagram that distinguish 
an improper ferroelectric from that of a proper ferroelectric$?$ In Figure~\ref{fig1}(a) 
we plot the energy as a function of the electric displacement field for YMnO$_3$. As 
expected from a ferroelectric compound, there are two symmetric potential wells, whose 
minima correspond to a spontaneous polarization of $P_{\rm s}=7.1$ $\mu$C/cm$^2$. 
This is slightly larger than previously reported values~\cite{Fennie/Rabe:2005,Sai_et_al:2009}; 
we ascribe the small difference to the different pseudopotentials used, 
and to the different (real-space) approach used here to compute the 
macroscopic polarization~\cite{Stengel/Spaldin:2005}.
Remarkably, the electrical equation of state is a multi-valued function in
the region near the origin.
%
Moreover, the curves belonging to the two symmetrical wells \emph{cross}
each other in $D=0$, rather than converging smoothly into a saddle point.
Both features are in stark contrast to the known behavior of ``standard'' ferroelectric 
materials that display a proper ferroelectric transition such as PbTiO$_3$ or BaTiO$_3$.

This is not completely unexpected, as the physical origin of ferroelectricity in 
YMnO$_3$ differs from that of BaTiO$_3$.
YMnO$_3$ is a classic improper ferroelectric, in which the spontaneous $P$ 
is driven by an unstable, non-polar ``trimerization'' ($K_3$) mode such that $P$ $\propto$ $K^3_3$.
Thus, rather than destabilizing the polarization, the $K_3$ distortion pushes the 
zone-center polar modes away from their centrosymmetric equilibrium configuration,
generating a polarization in the process.
In other words, the $K_3$ mode acts like a switchable 
``geometric field''~\cite{Fennie/Rabe:2005} that, depending on its orientation, 
produces two inequivalent (but related by a mirror symmetry operation) states 
of the system with opposite tendency towards a polar distortion.
Either state shows an almost perfectly linear dielectric behavior 
[see Figure~\ref{fig1}(b), the electric field follows an almost perfect straight 
line], with an average dielectric constant $\bar{\epsilon}_{33}\sim 13$, and an 
extremely weak piezoelectric response, of the order of $d_{33}\sim 1$ pm/V.
Note that the amplitude of the K$_3$ phonon mode does not change much across
the whole range of $D$ values, max $\pm 5$ \% with respect to the ferroelectric
ground state.
This is a consequence of the relatively weak coupling (only third-order) between 
the K$_3$ mode and $P$, and to the comparatively high energy associated
to the K$_3$ structural transition, which is more than three times higher than
the highest electrostatic energy, about 0.15 eV, obtained in our simulations.
%
In passing, we note that a similar kink in the electrical equation of
state at $P=0$ was previously reported for another ferroelectric material, 
KNO$_3$, where there is also a structural mode that produces a bistable
state in the origin~\cite{Dieguez/Vanderbilt:2006,Dieguez_KNO3}. 
We identify this characteristic kink, which is a consequence of the above-mentioned 
``geometric field'', 
as the \emph{ab-initio} fingerprint of improper ferroelectricity.



\begin{table}
\begin{ruledtabular}
\begin{tabular}{ccccc}
                 & FE$_{z}$ & AFD$_{\rm zi}$ & AFD$_{\rm zo}$ & (AFD/FE)$_z$ \\
\hline
$P_z$            &  30      &  0             &   0             &    32        \\
$\phi_{\rm zo}$ &   0      &  0             &   5.4           &    4.6       \\
$\phi_{\rm zi}$ &   0      &  3.3           &   0             &    3.1       \\
$\Delta E$      &  -12     &   -4           &   -32           &    -56       \\
\end{tabular}
\caption{ \label{tab1} Properties of the relevant phases in PTO/STO. The spontaneous
polarization $P_z$ is in $\mu$C/cm$^2$; the angles $\phi$ are in degrees; energies
$\Delta E$ (referred to the $P4/mmm$ phase) are in meV.}
\end{ruledtabular}
\end{table}


Our calculations indeed show, in agreement with Ref.~\onlinecite{Sai_et_al:2009}  
that there is a non-zero polarization in open circuit, $P_{\rm OC}$, 
which corresponds to the $D=0$ point in the electrical diagrams of 
Figure~\ref{fig1}.
Since $D=0$, $P_{\rm OC}=-\mathcal{E}_{\rm OC}/4\pi$, where $\mathcal{E}_{\rm OC}=-0.62$ GV/m 
is the internal field; this yields $P_{\rm OC}= \pm 0.55$ $\mu$C/cm$^2$. 
(The ratio between the spontaneous polarizations in short circuit and open
circuit corresponds to the average dielectric constant of YMnO$_3$ within
this region of the electrical diagram: 
$P_{\rm s}/P_{\rm OC}=\bar{\epsilon}_{33}\sim$13.)
Note that these values depart significantly from those 
($P_{\rm OC}=$6.1 $\mu$C/cm$^2$, $\mathcal{E}_{\rm OC}=$ 0.25 GV/m) reported in 
Ref.~\onlinecite{Sai_et_al:2009}.
To account for this discrepancy, we can speculate that the unsupported films of 
Ref.~\onlinecite{Sai_et_al:2009} are not under ideal open-circuit boundary conditions: 
The metallic oxygen-terminated surfaces would act effectively as electrodes, that 
partially screen the polarization charges of YMnO$_3$.
This highlights the importance of the analysis introduced in this work for understanding 
the intrinsic properties of improper ferroelectrics.
We shall come back to this point in Section~\ref{sec:capa}, where we show how one can
predict, from the calculated bulk data alone, the behavior of a film under \emph{any} 
type of electrical boundary conditions, ranging from open-circuit to close-circuit, 
and including all intermediate ``imperfect screening''~\cite{Junquera/Ghosez:2003} 
regimes.
%

\subsection{PbTiO$_3$/SrTiO$_3$ superlattices}



\begin{figure}
\includegraphics[width=3.3in]{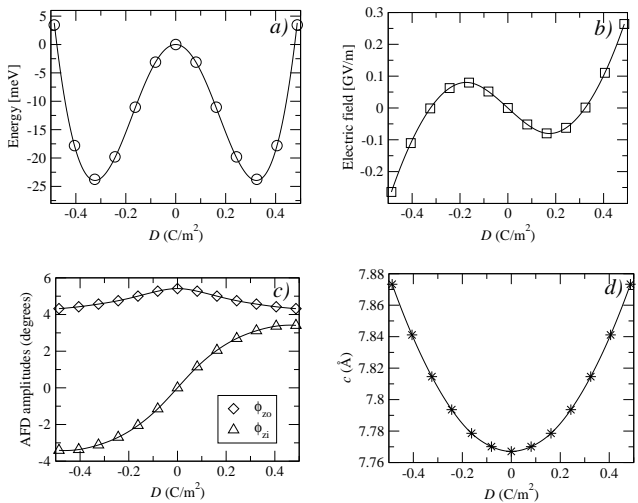}
\caption{Electrical properties of the PTO/STO superlattice. The energy (a), internal
electric field (b), AFD mode amplitudes (c) and out-of-plane strain (d) are plotted
as a function of the macroscopic $D$ field.}
\label{fig2}
\end{figure}

We now consider the PTO/STO superlattice.
In the high-symmetry $P4/mmm$ reference phase we find three unstable 
distortion modes, in agreement with Ref.~\onlinecite{Bousquet/Ghosez:2008}; a zone-center polar mode, plus two symmetry-inequivalent AFD 
rotations of the oxygen octahedrals along the $z$ axis, either out of phase
(AFD$_{\rm zo}$) or in phase (AFD$_{\rm zi}$).
(Note that we restrict our analysis to modes with rotation/displacement axis
along $z$.)
By freezing in each of these instabilities separately and then performing full structural relaxations we obtain three new phases of lower symmetry labeled FE$_z$, AFD$_{zi}$, and AFD$_{zo}$ in Table~\ref{tab1}. Furthermore, we obtain a fourth phase,  (AFD/FE)$_z$, in which all three instabilities coexist.
While we find some discrepancies at the quantitative level between our results and that of Ref.~\onlinecite{Bousquet/Ghosez:2008}, the level of qualitative agreement confirms their main conclusion.
%
%
The energy gains associated to ferroelectric transitions [$P4/mmm \rightarrow$ FE$_z$, AFD$_{\rm zo} \rightarrow$(AFD/FE)$_z$] are nevertheless systematically larger by a factor of 3-4 here  (respectively -3 and -8 meV in 
Ref.~\onlinecite{Bousquet/Ghosez:2008}, to be compared with -12 and -56+32=-24 meV in Table~\ref{tab1}), while the AFD$_{zi}$ and AFD$_{zo}$ instabilities appear systematically less pronounced (-28 and -73 meV in 
Ref.~\onlinecite{Bousquet/Ghosez:2008}). 
%
Ferroelectric perovskites are known to be especially sensitive to the computational approximations and these differences can be traced back in the different pseudopotentials used in both calculations and in the subsequent different in-plane lattice constant at which the calculations have been performed.

The large energy gains associated with the ferroelectric transition that we calculate suggests in fact that the polarization might be one of the primary order parameters (the other being AFD$_{\rm zo}$) rather than the two rotation distortions. Applying the approach outlined in this work should clarify this point. In Figure~\ref{fig2} we summarize the non-linear electrical and structural 
response of the superlattice as a function of $D$, analogous plots as in Figure~\ref{fig1}. 
Unlike YMnO$_3$, PTO/STO appears to behave in
all respects like a standard ferroelectric material. The double-well potential
in Figure~\ref{fig2}(a) is a single-valued function, differentiable at the 
saddle point, qualitatively similar to typical soft-mode driven systems like 
PbTiO$_3$ and BaTiO$_3$. 
This was to some extent unexpected : recent reports of improper ferroelectricity
on this same system suggested that its properties might closely mirror those of 
YMnO$_3$~\cite{Bousquet/Ghosez:2008}.
%
\begin{figure}
\includegraphics[width=3.5in,clip]{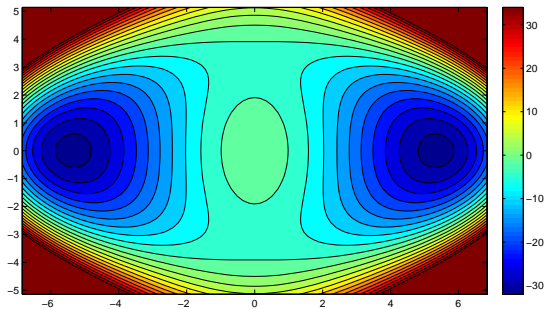}
\caption{Two-dimensional energy landscape of the AFD modes in the
PTO/STO superlattice around the $P4/mmm$ reference structure. Horizontal 
($\phi_{zo}$) and vertical ($\phi_{zi}$) axes are the mode amplitudes in 
degrees. The unit of the color bar is meV.}
\label{fig3}
\end{figure}
To gain insight into our apparently contradictory results, it is useful to look 
at the response of the AFD modes  -- which interact nontrivially with the 
ferroelectric polarization -- to an applied $D$-field.
In Figure~\ref{fig2}(c) we show the amplitude of the out-of-phase AFD$_{\rm zo}$ 
and of the in-phase AFD$_{\rm zi}$ modes as a function of $D$. 
At $D=0$, only a pure AFD$_{\rm zo}$ distortion is present (i.e., this configuration 
is identical to the pure AFD$_{zo}$ phase reported in Table~\ref{tab1}), while the 
AFD$_{\rm zi}$ distortion and the polarization are absent (AFD$_{zo}$ is a 
centrosymmetric space group.)  
This explains why we do not observe a linear behavior of the energy around 
$D=0$ in Figure~\ref{fig2}(a).
Only when we move away from $D=0$ do we see a linear increase of AFD$_{\rm zi}$,
the signature of the tri-linear coupling discussed in Ref.~\onlinecite{Bousquet/Ghosez:2008}. 

To illustrate why the AFD$_{\rm zo}$ and AFD$_{\rm zi}$ do not coexist
at $D=0$, we performed a series of calculation where we freeze in
various amplitudes of either mode, starting from the relaxed superlattice 
in the reference $P4/mmm$ symmetry.
In Figure~\ref{fig3} we plot the energy as a function of the two independent 
amplitudes $(\phi_{\rm zo},\phi_{\rm zi})$ within the restricted two-dimensional 
subspace spanned by these modes. As speculated above, both
modes are unstable in the reference structure, but the condensation of 
either mode quickly suppresses the instability with respect to the other 
(consistent with the results of Ref.~\cite{Bousquet/Ghosez:2008}). 
%
%
The global minima of the two-dimensional energy landscape corresponds to
pure AFD$_{\rm zo}$ states (the out-of-phase instability is much stronger
than the in-phase one), in agreement with the $D=0$ ground state found in
our electrical diagram of Figure~\ref{fig2}(a).
%
%


In the following Section we further substantiate these points by comparing 
the performance of hypothetical PTO/STO- or YMnO$_3$-based capacitors with 
real electrodes (i.e. under imperfect screening conditions).

\subsection{From bulk to thin-film capacitors}

\label{sec:capa}

\begin{figure}
\includegraphics[width=3.2in]{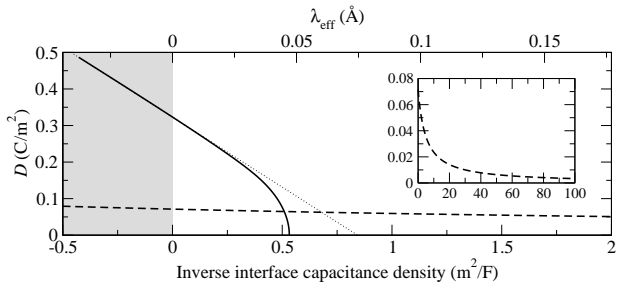}
\caption{Equilibrium $D$ for a thin-film PTO/STO (solid line) or
YMnO$_3$ (dashed line) capacitor as a function
of $\lambda_{\rm eff}/N$ or $\mathcal{C}^{-1}_{\rm I}/N$. 
The shaded area refers to a hypothetical overscreening regime with 
$\lambda_{\rm eff} < 0$.
The dotted line is a guide to the eye, to highlight the linear behavior
of $D$ in PTO/STO for $\lambda_{\rm eff}/N < 0.025$ \AA. The inset shows 
the absence of ``critical thickness'' in YMnO$_3$ in the limit of
large $\lambda_{\rm eff}/N$.}
\label{fig4}
\end{figure}

So far we have only discussed the \emph{bulk} electrical properties of YMnO$_3$ and
short-period PTO/STO superlattices. However, for practical applications in
thin-film capacitors, it is necessary to take possible finite size effects
into account.
In particular, the imperfect screening at a realistic film/electrode 
interface is known to affect the stability of the ferroelectric state by 
altering the electrical boundary conditions of the 
film~\cite{nature_mat,Junquera/Ghosez:2003}.
%
In the following we shall demonstrate that the impact of such interface effects 
can be directly predicted from the bulk equations of state of Fig.~\ref{fig1} 
and \ref{fig2}, i.e. without doing explicitly a demanding calculation of an 
actual capacitor.

We assume two symmetric electrode/ferroelectric interfaces, which behave as  
two thin layers of linear dielectric interposed between the insulator and the 
metal. These can be characterized indifferently by an interface capacitance
density per unit area $\mathcal{C}_{\rm I}$, or by an effective screening 
length  $\lambda_{\rm eff} = \epsilon_0 / \mathcal{C}_{\rm I}$~\cite{nature_mat}.
Then, the electrical equation of state of such a device can be written as~\cite{Stengel-09.2}
\begin{equation}
U(D) = N U_{\rm Bulk}(D) + 2\frac{SD^2}{2 \mathcal{C}_{\rm I}},
\end{equation}
where $U_{\rm Bulk}$ is the energy as in Fig.~\ref{fig1}(a) and \ref{fig2}(a);
$S$ is the cell cross-section and $N$ is the thickness of the film expressed 
in number of bulk units.
The minimum of $U(D)$ within short-circuit boundary conditions (zero applied
potential) is given by the stationary point $dU/dD=0$ (recall that the 
potential $V$ is related to $U$ by $SV(D)=dU/dD$~\cite{Stengel-09.2}),
\begin{equation}
V(D) = N V_{\rm Bulk}(D) + \frac{2D}{\mathcal{C}_{\rm I}} = 0.
\end{equation}
This can be written more compactly in terms of the interfacial 
``effective screening length'' $\lambda_{\rm eff} = \epsilon_0 / \mathcal{C}_{\rm I}$ as
\begin{equation}
\frac{\lambda_{\rm eff}}{\epsilon_0 N} =  -\frac{V_{\rm Bulk}(D)}{2D}.
\label{eqlambda}
\end{equation}
By inverting Eq.~\ref{eqlambda} we can
predict the behavior of a thin film of arbitrary thickness $N$ and with 
an arbitrary $\lambda_{\rm eff}$ (the ground state depends on
$\lambda_{\rm eff}/N$ only).

In Fig.~\ref{fig4}(a) we plot the resulting equilibrium $D$ of a hypothetical
$N=1$ capacitor as a function of $\lambda_{\rm eff}$. (We also consider
small negative values of $\lambda_{\rm eff}/N$ to account for a possible
overscreening~\cite{nature_mat,Kopp/Mannhart:2009}.)
PTO/STO (solid line) shows the typical destabilization of
the ferroelectric state (critical thickness for ferroelectricity) under an
imperfect screening regime; this occurs at $\lambda_{\rm eff} / N$=0.047 \AA.
Conversely, YMnO$_3$ appears almost unsensitive to the electrical boundary conditions,
with a spontaneous polarization that persists up to arbitrarily large values
of $\lambda_{\rm eff}/N$.
The limit $\lambda_{\rm eff}/N \rightarrow \infty$ corresponds to open-circuit
boundary conditions, where $D$ tends to zero asymptotically as $\sim N/\lambda_{\rm eff}$.
%

\subsection{From first principles to Landau theory}

While our first-principles results yield a fairly detailed picture on
the electrical behavior of the compounds under consideration, it is 
very useful to recast these data in terms of a simplified physical
model, where the role played by the active degrees of freedom emerges
clearly.
Also, most researchers in the \emph{ab-initio} community are used to
analyzing the behavior of ferroelectrics by performing calculations
under \emph{zero} external field, where the energy landscape as function of 
the active structural distortions is explored by freezing in various
amplitudes of selected imaginary-frequency phonons.
In this Section we shall trace a formal link between the constrained-$D$
method used in this work and the more familiar understanding of ferroelectrics, 
within either Landau theory or the frozen-phonon approach.

Consider the zero-field first-principles Kohn-Sham total energy, 
written as a function of the electronic degrees of freedom $\psi$ and the 
ionic positions $u$ (we omit the explicit discussion of the strain to avoid
complicating the notation), $E_{\rm KS}(\{\psi\},\{u\})$.
Around the high-symmetry configuration, one can construct a harmonic 
expansion in terms of the $N$ lattice eigenmode amplitudes 
$\xi_i$, 
\begin{equation}
E_{\rm KS}(\{ \xi \}) \sim \alpha_1 \xi_1^2 + \ldots + \alpha_N \xi_N^2,
\end{equation}
with the assumption that all $\xi_i$ vanish in the reference structure.
%
The polarity of the phonons can be expressed via the mode effective 
charge, $Z_i = \Omega \, dP/d\xi_i$ (we assume transversal zone-center modes, and we restrict
the analysis to one component of the polarization only).
Assume that the lattice distortions do not significantly affect the
purely electronic dielectric constant, $\epsilon_\infty$, which we
consider \emph{linear} in the applied external field; then
we can readily write the constrained-$D$ functional 
\begin{equation}
U(\{ \xi \}, D) = E_{\rm KS}(\{ \xi \})  + 
\frac{\Omega}{2 \epsilon_0 \epsilon_\infty}(D - P_\xi)^2,
\end{equation}
where $P_\xi$ is the lattice-mediated contribution to the polarization,
\begin{equation}
P_\xi = \frac{1}{\Omega} (\xi_1 Z_1 + \ldots \xi_N Z_N).
\end{equation} 
The ground state at a given $D$ is then obtained by taking the global
minimum with respect to all degrees of freedom,
\begin{equation}
U(D) = \min_{\xi_i} U(\{ \xi \}, D)
\end{equation}

Now, to make progress from here it is useful to note that not all 
phonon modes are equally important to describe the material. 
Most modes are hard, with little polar activity. Only
few low-energy phonons display a marked anharmonic behavior or are
unstable; these are the modes that we wish to focus our attention on. 
Typically, the hard modes are discarded from the Hamiltonian, or
better, are incorporated into an effective background constant; 
conversely, the low-energy modes are kept explicit, and higher-order 
terms are introduced in the Hamiltonian to ensure their proper
description.
%
Assume for example that only one mode, $\xi_s$, is ``soft'', and 
that it can be reasonably well described by a quadratic plus a quartic term,
\begin{equation}
E_{\rm KS}(\xi_s) = \alpha_s \xi_s^2 +\beta \xi_s^4 .
\end{equation}
Assume also that the phonon-phonon interaction is small, and that the electronic 
susceptibility is not significantly affected by $\xi_s$. Then the constrained-$D$
functional can be simplified by taking the minimum with respect to the hard modes 
only,
\begin{eqnarray}
U(\xi_s,D) &=& \min_{\xi_i, i\ne s} U(\{ \xi \}, D) \nonumber\\ 
           &=& \alpha_s \xi_s^2 +\beta \xi_s^4 + 
\frac{\Omega}{2 \epsilon_0 \epsilon_{\rm B}}(D - P_s)^2.
\end{eqnarray}
Here we have introduced the \emph{background dielectric constant},~\cite{Tagantsev:1986}
$\epsilon_{\rm B}$, which is given in terms of $\epsilon_\infty$,
the (hard) mode charges $Z_i$ and their eigenfrequencies $\omega_i$,
\begin{equation}
\epsilon_{\rm B} = \epsilon_\infty + \sum_{i \ne s} \frac{1}{\Omega} \frac{Z_i^2}{\omega_i^2}.
\end{equation}
Open-circuit boundary conditions correspond to setting $D=0$ 
in the above equation. In such a regime it is easy to see that
the electrostatic energy term in $U$ is just an additional 
quadratic (and positive) term in the soft mode amplitude,
of the type $\Delta \xi_s^2$.
Typically, this additional term completely suppresses the
ferroelectric instability: $\alpha_s + \Delta > 0$.

The above discussion establishes an important result: if we know the
approximate low-energy Hamiltonian in zero field, the ``background 
permittivity'' and the effective charges of the active modes (all these 
ingredients can be indifferently extracted from \emph{ab initio} 
calculations or from a phenomenological analysis) we can immediately 
calculate the ground state at a specific value of $D$. In other
words, the so-called constrained-$D$ method is by no means specific
to \emph{ab initio} calculations, and can be applied just as well 
to effective Hamiltonians and phenomenological descriptions. 
Indeed, mapping the properties of a given material in 
``electric displacement space''~\cite{Hong/Vanderbilt:11} may be a powerful tool 
to validate second-principles or macroscopic approaches against
the microscopic first-principles theory -- if the descriptions are 
of the same quality, the results should coincide.
In the following we discuss how these ideas apply to our two
testcases, YMnO$_3$ and PTO/STO.

In YMnO$_3$ the low-energy Hamiltonian can be written as a function
of a polar (but \emph{not} soft) $\Gamma$ mode and by an unstable
$K$-point mode,~\cite{Fennie/Rabe:2005}
\begin{eqnarray}
E_{\rm KS}(\xi_\Gamma,\xi_K) &\sim& \alpha_\Gamma \xi_\Gamma^2 + 
\alpha_K \xi_K^2 + \nonumber\\
&+& \beta_{KK} \xi_K^4 + \beta_{\Gamma K} \xi_\Gamma^2 \xi_K^2 + 
\gamma \xi_\Gamma \xi_K^3
\end{eqnarray}
The electrical variables have little impact on the structural $K$-point
mode, that acts like a ``switch'' (see Fig.~\ref{fig1}), and can therefore
be approximated by a constant with $\pm$ sign,
\begin{equation}
E_{\rm KS}(\xi_\Gamma) \sim \alpha_\Gamma \xi_\Gamma^2 + 
\tilde{\beta} \xi_\Gamma^2  \pm \tilde{\gamma} \xi_\Gamma.
\end{equation}
Typical values of the parameters have been reported by Fennie and Rabe
~\cite{Fennie/Rabe:2005}. The biquadratic coupling renormalizes the 
$\Gamma$-mode frequency by further \emph{hardening} it -- the new 
quadratic coefficient is given by $\tilde{\alpha} =\alpha_\Gamma + \tilde{\beta}$.
The last term is responsible for the geometric field.
If we plug this Hamiltonian into the electrostatic energy functional
we obtain
\begin{equation}
U(\xi_\Gamma,D) = \tilde{\alpha} \xi_\Gamma^2 \pm \tilde{\gamma} \xi_\Gamma +
\frac{\Omega}{2 \epsilon_0 \epsilon_{\rm B}}(D - P_\Gamma)^2,
\end{equation}
where $P_\Gamma = Z_\Gamma \xi_\Gamma / \Omega$. This is a second-order polynomial 
that can be solved analytically for the ground state at a given $D$,
\begin{equation}
U(D) = \frac{\Omega}{2 \epsilon_0 \epsilon_{\rm TOT}} (D \pm P_0)^2
\end{equation}
where $\epsilon_{\rm TOT}$ is the \emph{total} dielectric
constant, including the contribution from the $\Gamma$ mode,
and $P_0$ is the spontaneous polarization of YMnO$_3$ in zero 
field.
This equation accurately describes the results shown in Fig.~\ref{fig1}, 
where YMnO$_3$ emerges from the calculations as a linear dielectric 
with $\epsilon_{\rm TOT}\sim 13$ and a switchable polarization 
$P_0 = 7.1$ C/m$^2$.
In particular, if we impose open-circuit boundary conditions by
setting $D=0$, the electric field is 
\begin{equation}
\mathcal{E}_{\rm OC} = \frac{dU}{dD}\Big|_{D=0} = \pm \frac{P_0}{\epsilon_0 \epsilon_{\rm TOT}}
\end{equation}
and the polarization is $P_{\rm OC} = -\epsilon_0 \mathcal{E}_{\rm OC} = \mp P_0 / \epsilon_{\rm TOT}$,
consistent with our analysis of the first-principles data. 

The case of PTO/STO is slightly more complicated, due to the presence 
of three active modes instead of two, but we can follow a similar line
of reasoning.
The basic physics can be understood in terms of three phonon modes 
of the high-symmetry $P4/mmm$ reference structure: a polar $\Gamma$
mode $\xi_\Gamma$, and two zone-boundary structural modes that 
correspond to out-of-phase and in-phase rotations of the oxygen
octahedra with respect to the $z$ axis, $\phi_{zo}$ and $\phi_{zi}$.
The energy in zero field can be expanded as
\begin{eqnarray}
E_{\rm KS}(\xi_\Gamma,\phi_{zo},\phi_{zi}) &=& A_\Gamma \xi_\Gamma^2 + B_{\Gamma \Gamma} \xi_\Gamma^4 + \nonumber\\
&+& A_{o} \phi_{zo}^2 + B_{oo} \phi_{zo}^4 + \nonumber\\
&+& A_{i} \phi_{zi}^2 + B_{ii} \phi_{zi}^4 + \nonumber\\
&+& B_{o\Gamma} \phi_{zo}^2 \xi_\Gamma^2 + B_{i\Gamma} \phi_{zi}^2 \xi_\Gamma^2 +
B_{oi} \phi_{zo}^2 \phi_{zi}^2 +\nonumber\\
&+& C \phi_{zo} \phi_{zi} \xi_\Gamma.
\end{eqnarray}
The first three lines correspond to the double-well potential associated with
each individual mode ($\alpha_\Gamma,\alpha_{zo},\alpha_{zi} <0$); the fourth 
line provides the lowest-order biquadratic couplings responsible for the pair-wise 
competition or cooperation between the modes; the last line is the tri-linear 
coupling, that was previously identified as key to a novel type of improper 
ferroelectricity in this system.
As the rotations do not directly contribute to $P$ (their dynamical charge is zero), 
the simplified constrained-$D$ functional can be written as
\begin{equation}
U(\xi_\Gamma,\phi_{zo},\phi_{zi},D) = E_{\rm KS}(\xi_\Gamma,\phi_{zo},\phi_{zi}) + 
\frac{\Omega}{2 \epsilon_0 \epsilon_{\rm B}}(D - P_\Gamma)^2.
\end{equation}
At $D=0$, the electrostatic energy term reduces to $\gamma \xi_\Gamma^2$ with $\gamma$ 
a positive constant; this effectively renormalizes the $A_\Gamma$
coefficient to $\tilde{A}_\Gamma = A_\Gamma + \gamma$. Recall
that $A_\Gamma$ is negative -- in our \emph{ab initio} model 
PTO/STO has a ferroelectric instability even in absence of rotations.
Conversely, $\tilde{A}_\Gamma$ is \emph{positive} and large. Indeed, the
electrostatic term in $U$ induce a shift in the frequency of the zone-center
optical modes which become \emph{longitudinal} rather than transverse,
and the LO-TO splitting in perovskite titanates is enormous.~\cite{Zhong-94}
As $\tilde{A}_\Gamma$ is positive, the only term 
in the Hamiltonian that can induce a polarization in open-circuit is precisely
the trilinear one $C \phi_{zo} \phi_{zi} \xi_\Gamma$. 
Furthermore, as $\tilde{A}_\Gamma$ is large, $\xi_\Gamma$
will be virtually \emph{suppressed} from the low-energy Hamiltonian,
and only a simultaneous presence of $zi$ and $zo$ rotations (at $\xi_\Gamma=0$)
will produce a geometric field.
\begin{table}
\begin{ruledtabular}
\begin{tabular}{ccccc}
    $A_o$   & $A_i$ & $B_{oo}$ & $B_{ii}$  & $B_{io}$ \\
\hline
  -2.1097        & -0.6912      &    0.0368      &    0.0363    &  0.1372      \\
\end{tabular}
\caption{ \label{supp:tab} Parameters of the model described in the
text. Values of the parameters are in eV. The independent variables 
$\phi$ are in degrees.}
\end{ruledtabular}
\end{table}
In order to verify this possibility, we need to study the energy as a function 
of the AFD modes amplitude, i.e. the energy landscape that is sketched in Fig.~3. 

In absence of $\xi_\Gamma$, the low-energy Hamiltonian reduces to
$$E_{\rm KS}(\phi_{zo},\phi_{zi}) = A_o \phi_{zo}^2 + A_i \phi_{zi}^2 +$$
\begin{equation}
 + B_{oo} \phi_{zo}^4 +
 B_{ii} \phi_{zi}^4 + B_{io} \phi_{zo}^2 \phi_{zi}^2.
\end{equation}
The values of the five parameters are reported in Table~\ref{supp:tab}; 
the maximum deviations with the actual first-principles data are lower
then 1.5 meV across the whole two-dimensional energy landscape.

The necessary and sufficient condition for having four minima where 
$\phi_{zo}$ and $\phi_{zi}$ are nonzero is that the ``pure'' minima
(where only one degree of freedom is active at the time) are in
fact saddle points of the two-dimensional function $E(\phi_{zo},\phi_{zi})$,
\begin{eqnarray}
A_o - \frac{A_i B_{io}}{2B_{ii}} &<& 0 \\
A_i - \frac{A_o B_{io}}{2B_{oo}} &<& 0 
\end{eqnarray}
It is easy to see that the first of these conditions is satisfied,
i.e. the AFD$_{\rm zi}$ configuration is not a true minimum but a 
saddle point in this reduced configuration space. 
Conversely, the second condition is violated, i.e. the pure AFD$_{\rm zo}$
state is indeed a global minimum, where the original in-phase AFD instability 
is suppressed.

\section{Discussion}

In the previous section we have clarified the physical ingredients that
are responsible for the calculated electrical behavior of YMnO$_3$ 
and PTO/STO. However, a substantial limitation of our work so
far has been the complete neglect of temperature effects. As the 
concept of ``improper ferroelectric'' is intimately bound to the
thermodynamics of phase transitions, it is useful to discuss here whether
we can infer a number of likely scenarios based on our zero-temperature
results. (Of course, our arguments will remain limited to a qualitative
level, as presently we don't have access to finite-temperature simulations.)
This point is particularly important in order to reconcile our 
findings (that a true ``geometric field''~\cite{Fennie/Rabe:2005} is not present 
in PTO/STO) with the experimental data of Ref.~\onlinecite{Bousquet/Ghosez:2008},
where the typical signatures of improper ferroelectricity emerge clearly.
Given the relevance of this system in the development of novel
multifunctional materials, we shall focus the following discussion
on PTO/STO.

Experimentally, PTO/STO undergoes a phase transition at $T_C=500$ K 
from a high-temperature paraelectric state to a polar state
where the spontaneous $P$ that increases linearly with decreasing $T$.
To rationalize this phase transition, several scenarios are possible,
depending on the respective behavior of the three active degrees of freedom.
First, we can imagine a transition where the AFD modes act as primary
order parameters in the transition, and the polarization emerges as
a secondary effect; this is the classic picture of improper ferroelectricity.
This, however, does not appear consistent with our simulations. In
absence of polarization, the two AFD modes are mutually exclusive, 
and they don't break inversion symmetry alone. In other words, 
if we suppress $P$ there is no ``geometric field'' in the system.
Second, we can imagine a transition where $\xi_\Gamma$ is the primary
order parameter, and the AFD distortions are secondary. This is,
however, inconsistent with the experiments of Ref.~\onlinecite{Bousquet/Ghosez:2008}:
in a second-order phase transition involving only $P$, one would
expect $P \propto \sqrt{T_C-T}$, which differs from the reported linear
behavior.

Interestingly, a third possibility was proposed in a recent work,~\cite{Etxebarria-10}
where analogous Hamiltonians involving a trilinear coupling were 
studied in a phenomenological framework. 
The scenario described in Ref.~\onlinecite{Etxebarria-10} is that of 
an ``avalanche'' first-order transition, where all three degrees 
of freedom condense simultaneously without any of them becoming
unstable. This is a situation where the usual assumptions of Landau
theory, and in particular the distinction between primary and 
secondary order parameters, break down: neither the polarization
nor the AFD modes would act as either master or slave.
This possibility can be simply understood by constructing a 
``reaction coordinate'' $Q$ and by imposing that all three 
degrees of freedom must be proportional to $Q$. Then
the Hamiltonian reduces to the following expression
\begin{equation}
E_{\rm KS}(Q) = \alpha Q^2 + \beta Q^4 + \gamma Q^3.
\end{equation}
All the degrees of freedom are individually stable above the
transition, so $\alpha$ is positive, and its magnitude will 
decrease with $T$. $\beta$ is necessarily positive at all
temperatures, or thermodynamic stability is lost. One can
easily see that if $\gamma$ is different from zero a 
secondary minimum will appear for sufficiently low $T$
(small $\alpha$), and will drop below the main minimum at $Q=0$
\emph{before} the $Q=0$ minimum becomes a saddle point.
Thus, in presence of this trilinear term in the Hamiltonian,
the transition is very likely to be first-order and involve all
degrees of freedom simultaneously.

Strictly speaking, even this hypothesis does not constitute an
ideal match to the experimental results. A first-order phase
transition necessarily involves a jump in the polarization
from zero to a finite value at $T_C$, while a continuous curve
was reported experimentally. 
One can imagine, however, that the transition might be only 
weakly first-order, and the discontinuity might be too 
small to be detectable in the experimental setup of Ref.~\onlinecite{Bousquet/Ghosez:2008}.
Another possibility is that the depolarizing field produced
by the electrodes might have an impact on the measured $P$,
or more generally that other degrees of freedom not
considered here (e.g. nanodomains) might play a role in the 
transition.
A last alternative is that standard DFT calculations do not properly
describe the relative strength of the different instabilities although
these methods have been successfully applied to many cases so far.

Resolving these issues goes beyond the scopes of the present
work; what we want to stress here is that the electrical analysis
used in this work constitutes a powerful tool to study complex
ferroelectric systems where many degrees of freedom interact
nontrivially with the polarization.

\section{Conclusions}

In conclusion, we have shown that constrained-$D$ calculations provide a powerful method 
to understand the electrical behavior of multifunctional oxide compounds. As practical
examples, we discussed two different material systems: A prototypical improper 
ferroelectric, YMnO$_3$, 
and short-period PTO/STO superlattices.
%
Concerning YMnO$_3$, our results have evidenced a characteristic ``kink'' in 
the electrical equation of state (energy $U$ as a function of the electric displacement $D$), 
which is responsible for the remarkable insensitivity of this material to the electrical 
boundary conditions.~\cite{Sai_et_al:2009}
Concerning PTO/STO, our results indicate that the electrical properties 
depends crucially on the interaction between the two non-polar AFD modes:
Only if the AFD modes are the dominant instabilities \emph{and} they do not 
compete with each other we expect the electrical properties to be qualitatively
similar to those of YMnO$_3$. While our PTO/STO model does not appear to satisfy 
either condition, recent results~\cite{Benedek/Fennie:2010} give strong indications 
for such a scenario to occur in a related material, Ca$_3$Mn$_2$O$_7$.
To illustrate the consequences of these findings in a realistic capacitor configuration, we showed 
how the $U(D)$ curve can be readily converted into a $P(\lambda_{eff}/N)$ curve, where $\lambda_{eff}$ is 
the effective screening length of the electrode interface and $N$ is the thickness.

As an outlook, Ref.~\onlinecite{Bousquet/Ghosez:2008} has introduced us to the remarkable possibility that two 
AFD rotation distortions can induce ferroelectricity via a trilinear coupling with the polarization.
This idea holds much promise to realize novel multifunctional phenomena, and has led to the 
discovery of an entirely new class of ``hybrid'' improper ferroelectrics~\cite{benedek12}.
While rationalizing these breakthroughs by putting them on firm theoretical grounds, our work also
illustrates some interesting new aspects of these systems that might be useful for future applications.
For example, it appears clear from our data that, in contrast to YMnO3, in PTO/STO the
AFD modes are very strongly coupled to the polarization.
This implies that the control of octahedral rotations by means of an applied field is
indeed possible, and the constrained-$D$ method appears as the ideal theoretical tool
to investigate (and engineer) this functionality in a variety of oxide systems.

%

\section*{Acknowledgments}

We acknowledge discussions with  K.M$.$ Rabe and J. \'I\~niguez. 
This work was supported by the DOE-BES under Award Number DE-SCOO02334 (CJF);
by DGI-Spain through Grants No. MAT2010-18113 and No. CSD2007-00041 (MS); 
by the European Commission through the project EC-FP7,
Grant No. NMP3-SL-2009-228989 ``OxIDes'' (PhG and MS); and by
the Francqui Foundation through a Research Professorship (PhG).
We thankfully acknowledge the computer resources, 
technical expertise and assistance provided by the 
Red Espa\~nola de Supercomputaci\'on (RES) and by the 
Supercomputing Center of Galicia (CESGA).

\bibliography{improper}
\end{document}